\documentclass{article}

\usepackage{amsmath,amssymb,amsthm}
\usepackage{dsfont,array}

\usepackage{times}

\usepackage{url}
\usepackage{xcolor}
\usepackage{listings}

\theoremstyle{definition}
\newtheorem{example}{Example}

\newcommand{\SETmemberof}{\mathsf{member\_of}}
\newcommand{\SETinsert}{\mathsf{insert}}

\newcommand{\logictrue}{\mathit{True}}

\newcommand{\pre}{\mathbf{pre}}
\newcommand{\post}{\mathbf{post}}
\newcommand{\ldoubsq}{[\![}
\newcommand{\rdoubsq}{]\!]}
\def\numset#1{\mathds{#1}}

\newcommand{\naturals}{\numset{N}}

\newcommand{\modeleq}{\asymp}

\begin{document}

\title{Specifying Reusable Components}

\author{Nadia Polikarpova
\and Carlo A. Furia
\and Bertrand Meyer
}

\date{}

\maketitle

\begin{abstract}
Reusable software components need well-defined interfaces, rigorously and completely documented features, and a design amenable both to reuse and to formal verification; all these requirements call for expressive specifications.
This paper outlines a rigorous foundation to \emph{model-based contracts}, a meth\-odology to equip classes with expressive contracts supporting the accurate design, implementation, and formal verification of reusable components.
Model-based contracts conservatively extend the classic Design by Contract by means of expressive models based on mathematical notions, which underpin the precise definitions of notions such as abstract equivalence and specification completeness.
Preliminary experiments applying model-based contracts to libraries of data structures demonstrate the versatility of the methodology and suggest that it can introduce rigorous notions, but still intuitive and natural to use in practice.
\end{abstract}


\section{Introduction} \label{sec:introduction}
The case for precise software specifications involves several
well-known arguments; in particular, specifications help understand
the problem before building a solution, and they are necessary for
verifying implementations. In the case of a library of reusable
software components, precise specifications have another application,
essential to the effective use of the library: providing client
programmers with a description of the interface (the API). To help
produce such specifications, Design by Contract techniques \cite{OOSC2} let
authors of reusable modules equip them with specification elements
known as ``contracts'' (routine preconditions and postconditions,
class invariants), which tools from the development environment can
extract to produce automatically generated API documentation.

While specifications primarily intended for purposes other than
component development typically use a specification language based on
mathematics, approaches using Design by Contract, such as Eiffel
\cite{OOSC2}, JML \cite{Leavens2005} and Spec\# \cite{Specsharp} rely
instead on an assertion language embedded in the programming
language. In Eiffel, for example, contracts are expressed through
assertions built out of the language’s Boolean expressions, with a few
extensions; the most notable of these extensions is the
\lstinline|old| notation which makes it possible to express
postconditions as properties of both the starting and ending states of
the computation. This approach adds a significant element to the list
of benefits of precise specifications: being expressed in the
programming language, contracts can be \emph{evaluated} during
execution. (We will use the term ``executable assertions'', although
this is really about evaluation rather than execution; another
possible term is ``embedded'' assertion, to emphasize that the
assertion language is included in the programming language.)  As a
consequence, contracts have played a major role in \emph{testing},
especially for Eiffel, where an advanced testing environment, AutoTest
\cite{Meyer2009}, takes advantage of contracts for automatic test
generation; more generally, Eiffel programmers routinely rely on
run-time contract evaluation for testing and debugging.

Another practical benefit of the approach is teachability: programmers
already understand Boolean expressions, and do not need to learn a
separate specification language.  These practical advantages of
executable assertions have traditionally come at a price:
expressiveness. Unlike a full-fledged specification language (such as
B \cite{Abrial1996}, based on set theory), an assertion language
embedded in a programming language makes it harder to express the full
specification of programs and components. As a typical example, the
postcondition of a ``push'' operation on a stack in the existing
standard Eiffel library expresses that the new top of the stack will
be the item just pushed, and that the number of items will have been
increased by one; but it typically does not state, except in the form
of a comment, that the \emph{other} elements of the stack are
unaffected. This example is typical: an extensive study
\cite{WritingContracts} indicates that in practice Eiffel classes
contain many contracts, but (see also \cite{Polikarpova2009}) they
cover only part of the programmer’s informal understanding of the
specification.

Can we retain all the advanced benefits of specifications, in
particular support completeness of specifications and static checks
(including proofs), while retaining an executable specification
language that can also be used for testing? The present work proposes
a positive answer, based on the idea of \emph{models}.

Specifications, in this approach, do not require any special language
beyond the classical assertion language embedded in the programming
language. Instead, they rely on a methodological principle: associate
with every class one or more \emph{model queries} specifying the
semantics of the associated objects through standard mathematical
concepts, represented by instances of \emph{model classes}. The model
classes are also expressed in the programming language, but they are
just direct translations of mathematical concepts (such as sets,
functions, relations etc.); they have no operational properties
(attributes (fields), assignment, side effects, procedures and such),
so that the corresponding objects are immutable. The model queries of
a normal (non-model) class are expressed in terms of such model
classes; for example a stack class can have a model query
\lstinline|sequence| of the model type \lstinline|SEQUENCE|,
associating a sequence with every stack (the sequence of stack items,
starting for example from the top). It is then possible to specify
operations of the class through their effect on the model queries; for
example the push operations yields a new stack whose
\lstinline|sequence| query yields a sequence starting with the element
being pushed and continuing with the elements of the original
sequence. In this example the class only has one model query
(sequence), but any number of model queries is possible; the model
queries can be existing features of the class, or new features added
for the sole purpose of specification.

This idea of \emph{model-based contracts} is not new; previous own work \cite{Schoeller2004,Schoeller2007} and, among others, JML \cite{Leavens2005} introduced the concept and provided libraries of model classes.
Developing a \emph{rigorous and systematic approach to model-based specifications} is the main contribution of the present paper.
Section \ref{sec:foundations} shows how the interface of a class defines unambiguously a notion of \emph{abstract space}, which in turn determines the model of the class; programmers can easily introduce model classes and model queries in accordance with this model.
Section \ref{sec:foundations} also outlines precise guidelines to write contracts that refer to the chosen model queries.
The guidelines come with a definition of \emph{completeness} of the postcondition of a feature with respect to the class model.
The definition is formal, yet amenable to informal reasoning; it is practically useful in assessing whether a contract is sufficiently detailed or is likely omitting some important details of what the feature achieves.

Section \ref{sec:mbc-at-work} describes two case studies that used this methodology for model-based specifications to develop libraries of data structures with strong contracts.
The results achieved show that the methodology is successful in delivering well-designed components with expressive --- usually complete --- specifications.
Most advantages of standard Design by Contract are retained, such as congeniality to programmers and ease of reasoning, while pushing a more accurate evaluation of design choices and an impeccable definition of interfaces.
The executability of most model classes even supports the reuse of Eiffel's automated contract-based testing infrastructure with more expressive contracts, which boosts the effectiveness of automated testing in finding defects in developed software.



\section{Motivation and overview}  \label{sec:overview}
Design by Contract (DbC) is a discipline of analysis, design, implementation, and management of software.
It relies on the fundamental idea of defining the role of any component in the system in terms of a \emph{contract} that formalizes the obligations and benefits of that component relative to the rest of the system.
Concretely, the contract is as a collection of assertions (\emph{preconditions}, \emph{postconditions}, and \emph{invariants}) that constitute the module's \emph{specification}.

\subsection{Some limitations of Design by Contract}
To emphasize the seamless connection that must exist between specification and implementation, and to make writing contracts palatable to the programmer, DbC uses the same notation for expressions in the implementation and in the specification.
This choice successfully encourages programmers to write contracts \cite{WritingContracts}.
On the other hand, it also restricts the assertions that can be expressed --- or that can be expressed easily.
This restriction ultimately impedes the formalization and verification of full functional correctness and even limits the scope of application of DbC for the correct design of an implementation.
Let us demonstrate this on a couple of examples from the EiffelBase library \cite{EiffelBase}.

Lines 1--14 in Table \ref{tab:linkedlistetal} show a portion of class \lstinline|LINKED_LIST|, implementing a dynamic list.
Features (members) \lstinline|count| and \lstinline|index| record respectively the number of elements stored in the list and the current position of the internal cursor.
Routine \lstinline|put_right| inserts an element \lstinline|v| to the right of the current position of the cursor, without moving it.
The postcondition of the routine (clause \lstinline|ensure|) asserts that inserting an element increments \lstinline|counter| by one but does not change \lstinline|index|.
This is correct, but it does not capture the gist of the semantics of insertion: the list after insertion is obtained by all the elements that were in the list up to position \lstinline|index|, followed by element \lstinline|v| and then by all elements that were to the right of \lstinline|index|.

Expressing such complex facts is impossible or exceedingly complicated with the standard assertion language; as a result most specifications are \emph{incomplete} in the sense that they fail to capture precisely the functional semantics of routines.
Weak specifications hinder formal verification in two ways.
First, establishing weak postconditions is simple, but confidence in the full functional correctness of a verified routine will be low: the quality of specifications limits the value of verification.
Second, weak contracts affect negatively verification modularity: it is impossible to establish what a routine $r$ achieves, if $r$ calls another routine $s$ whose contract is not strong enough to document its effect within $r$ precisely.

\setcounter{lstnumber}{2}
\lstset{basicstyle=\scriptsize}
\begin{table}[!tb]
\begin{tabular}{m{.43\textwidth} m{.57\textwidth}}
\begin{lstlisting}
class LINKED_LIST [G]
	count: INTEGER -- Number of elements
			
	index: INTEGER	-- Current cursor position

	put_right (v: G)
	  -- Add `v' to the right of cursor.
		require $\ $   0 <= index <= count
		do $\ \ldots$
		ensure
			count = old count + 1
			index = old index
		end
\end{lstlisting} \addtocounter{lstnumber}{1}
&
\begin{lstlisting}
   duplicate (n: INTEGER): LINKED_LIST
     -- Copy of sublist of length `n' beginning at current position
     require n >= 0  do $\ \ldots\ $  ensure Result.index = 0 $\ $ end
end

class TABLE [G, K]
	put (v: G ; k: K)
    -- Associate value `v' with key `k'.
    require valid_key (k)
    deferred	end

end
\end{lstlisting}
\end{tabular}
\caption{Snippets from the EiffelBase classes \lstinline|LINKED_LIST| (lines 1--17) and \lstinline|TABLE| (lines 19--25).}
\label{tab:linkedlistetal}
\end{table}
\lstset{basicstyle=\footnotesize}



Weak assertions limit the potential of many other applications of DbC.
Specifications, for example, should document the abstract semantics of operations in deferred classes (classes without an implementation).
Weak contracts cannot fully do so; as a result, programmers have fewer safeguards to prevent inconsistencies in the design and fewer chances to make deferred classes useful to clients through polymorphism and dynamic dispatching.

Feature \lstinline|put| in class \lstinline|TABLE| (lines 16--19 in Table \ref{tab:linkedlistetal}) is an example of such a phenomenon.
It is unclear how to express the abstract semantics of \lstinline|put| with standard contracts.
In particular, the absence of a postcondition leaves it undefined what should happen when an element is inserted with a key that is already associated to some other element: should \lstinline|put| replace the previous element with the new one or cancel the insertion of the new element?
Indeed, some heirs of \lstinline|TABLE| implement \lstinline|put| with a replacement semantics (such as class \lstinline|ARRAY|), while others disallow overriding of preexisting mappings with \lstinline|put| (such as class \lstinline|HASH_TABLE|).
Some classes (including \lstinline|HASH_TABLE|) even introduce another feature \lstinline|force| that implements the replacement semantics.
This obscures the behavior of routines to clients and makes it questionable whether \lstinline|put| has been introduced at the right point in the inheritance hierarchy.

\subsection{Enhancing Design by Contract with models}
This paper presents an extension of DbC that addresses the aforementioned problems.
The extension conservatively enhances DbC with \emph{model classes}: immutable classes representing mathematical concepts that provide for more expressive specifications.
Wrapping mathematical entities with classes supports richer contracts without need to extend the notation, which remains the one familiar to programmers as in DbC.
Contracts using model classes are called \emph{model-based contracts}. 

\setcounter{lstnumber}{2}
\lstset{basicstyle=\scriptsize}
\begin{table}[!tb]

\begin{tabular}{m{.56\textwidth} m{.45\textwidth}}
\begin{lstlisting}
note model: sequence, index
class LINKED_LIST [G]
   sequence: MML_SEQUENCE [G]
     -- Sequence of elements
     do $\ \ldots\ $ end

   count: INTEGER -- Number of elements
     ensure Result = sequence.count end
			
   index: INTEGER	-- Current cursor position

   put_right (v: G)
	  -- Add `v' to the right of cursor.
		require $\ $   0 <= index <= count
		do $\ \ldots$
		ensure 
       sequence = old ( sequence.front (index)
         .extended (v) + sequence.tail (index + 1) )
       index = old index
		end
end
\end{lstlisting} \addtocounter{lstnumber}{1}
&
\begin{lstlisting}
note model: map
class TABLE [G, K]
   map: MML_MAP [G, K]
    -- Map of keys to values
    deferred end		

   put (v: G ; k: K)
    -- Associate value `v' with key `k'.
    require map.domain [k]
    deferred
    ensure
     map = old map.replaced_at (k, v)
    end
end
\end{lstlisting}
\end{tabular}
\caption{Classes \lstinline|LINKED_LIST| (left) and \lstinline|TABLE| (right) with model-based contracts.}
\label{tab:mb-linkedlistetal}
\end{table}
\lstset{basicstyle=\footnotesize}

Table \ref{tab:mb-linkedlistetal} shows an extensions of the examples in Table \ref{tab:linkedlistetal} with model-based contracts.
\lstinline|LINKED_LIST| is augmented with a query \lstinline|sequence| that returns an instance of class \mbox{\lstinline|MML_SEQUENCE|,} a model class representing a mathematical sequence of elements of homogeneous type; the implementation, omitted for brevity, builds \lstinline|sequence| according to the actual content of the list.
The meta-annotation \lstinline|note| declares the two features \lstinline|sequence| and \lstinline|index| as \lstinline|model| of the class; every contract will rely on the abstraction they provide.
In particular, the postcondition of \lstinline|put_right| can precisely describe the effect of the routine: the new \lstinline|sequence| is the concatenation of the \lstinline|old sequence| up to \lstinline|index|, extended with element \lstinline|v|, with the tail of the \lstinline|old sequence| starting after \lstinline|index|.
We can assert that the new postcondition --- including the clause about \lstinline|index| --- is \emph{complete} with respect to the model of the class, because it completely defines the effect of \lstinline|put_right| on the abstract model.
This notion of completeness is a powerful guide to writing accurate specification that makes for well-defined interfaces and verifiable classes.

The mathematical notion of a \lstinline|map| --- encapsulated by the model class \lstinline|MML_MAP| --- is the natural model for the class \lstinline|TABLE|.
Feature \lstinline|map| cannot have an implementation yet, because \lstinline|TABLE| is deferred and hence it is not committed to any representation of data.
Nonetheless, the mere availability of a model class supports complex specifications already at this abstract level.
In particular, writing a complete postcondition for routine \lstinline|put| requires to commit to a specific semantics for insertion.
The example in Table \ref{tab:mb-linkedlistetal} chooses the replacement semantics; correspondingly, all heirs of \lstinline|TABLE| will have to conform to this semantics, guaranteeing a coherent reuse of \lstinline|TABLE| throughout the class hierarchy.


\section{Foundations of model-based contracts}\label{sec:foundations}

\subsection{Specifying classes with models} \label{sec:spec-class-with}
This subsection describes a rigorous approach to equipping classes with expressive contracts.

\subsubsection{Interfaces, references, and objects.}
The definitions of abstract objects and models (introduced in the remainder) rely on the following simple assumptions about classes.
A class $C$ denotes a collection of objects.
Expressions such as $o:C$ define $o$ as a reference to an object of class $C$; the notation is overloaded for conciseness, so that occurrences of $o$ can denote the object it references or the reference itself, according to the context.
Each class $C$ defines a notion of \emph{reference} equality $\equiv_C$ and of \emph{object} equality $\circeq_C$; 
both are equivalence relations. 
Two objects $o_1, o_2 : C$ of class $C$ can be \emph{reference equal} (written $o_1 \equiv_C o_2$) or \emph{object equal} (written $o_1 \circeq_C o_2$).
Reference equality is meant to capture whether $o_1$ and $o_2$ are aliases for the same physical object, 
whereas object equality is meant to hold for (possibly) physically distinct objects with the same actual content.
The following discussion is however independent of the particular choice of reference and object equality.

The principle of information hiding prescribes that each class define an interface: the set of its publicly accessible features \cite{OOSC2}. 
It is good practice to partition features into queries and commands; queries are functions of the object state, whereas commands modify the object state but do not return any value.
$I_C = Q_C \cup M_C$ denotes the interface of a class $C$ partitioned in queries $Q_C$ and commands $M_C$.\footnote{Constructors need no special treatment and can be modeled as queries returning new objects.}
It is convenient to partition all queries into \emph{value-bound} queries $Q_C^o$ and \emph{reference-bound} queries $Q_C^r$.
Value-bound queries should create fresh objects to return (or more generally objects that were unknown to the client before calling the query), whereas reference-bound queries give the client direct access, through a reference, to parts of the target object or of the query arguments.
In other words, clients of a value-bound query are insensitive to whether they received a unique fresh object or they are just sharing a reference to a previously existing one.
The chosen partitioning between value-bound and reference bound queries does not affect the following discussion, although it is usually quite natural to adhere to this informal distinction when designing a class.

\begin{example}
Query \lstinline|item| (Table \ref{tab:mb-linkedlistmore}) is reference-bound, 
as the client receives the very same physical object that was earlier inserted in the list. 
Query \lstinline|duplicate| (Table \ref{tab:mb-linkedlistmore}) is instead value-bound, as it returns a copy of a portion of the list.
\end{example}

The classification in value-bound and reference-bound extends naturally to \emph{arguments} of features: if the feature does not rely on having a direct reference to the actual argument (as opposed to a copy of it), the argument is value-bound; otherwise, it is reference-bound.

\subsubsection{Abstract object space.}
The interface $I_C$ induces an equivalence relation $\modeleq_C$ over objects of class $C$ called \emph{abstract equality} and defined as follows: 
$o_1 \modeleq_C o_2$ holds for $o_1, o_2 : C$ 
iff for any applicable sequence of calls to commands $m_1, m_2, \ldots \in M_C^*$ and a query $q \in Q_C$ returning objects of some class $T$, 
the qualified calls $o_1.m_1; o_1.m_2;\cdots$ and $o_2.m_1; o_2.m_2;\cdots$ (with identical actual arguments where appropriate) drive $o_1$ and $o_2$ in states such that if $q$ is reference-bound then $o_1.q \equiv_T o_2.q$, 
and if $q$ is value-bound then $o_1.q \circeq_T o_2.q$.
Intuitively, two objects are equivalent with respect to $\modeleq_C$ if a client cannot distinguish them by any sequence of calls to public features.

Abstract equality defines an \emph{abstract object space}: the quotient set $A_C = C/\modeleq_C$ of $C$ (as a set of objects) by $\modeleq_C$.
As a consequence, two objects are equivalent w.r.t.\ $\modeleq_C$ iff they have the same \emph{abstract (object) state}.
Any concrete set that is isomorphic to $A_C$ is called a \emph{model} of $C$.

\begin{example}
\label{ex:queues}
A \emph{queue} class typically consists of the queries \lstinline|item|, \lstinline|count|, and \lstinline|empty| --- returning the next element to be dequeued, the total number of elements in the queue, and a fresh empty queue --- and the commands \lstinline|put| and \lstinline|remove| --- to enqueue an element and dequeue the next element.
If \lstinline|remove| were not part of the interface, any element in the queue but the least recently inserted one would be inaccessible to clients; the model of such a class would then be a pair of type $\naturals \times G$ recording the current number of elements and the latest enqueued element of generic type $G$.
Including \lstinline|remove| in the interface, as it usually is the case for queues, allows clients to read the whole sequence of enqueued elements.
Hence, two queues with full interfaces are indistinguishable iff they have the very same sequence of elements; the model of a queue class with full interface is then an abstract sequence of type $G^*$.
\end{example}

As all the following examples will suggest, the most natural design choice implements object equality to have the same semantics as abstract equality.
Notice, however, that complying or not with this rule of thumb does not affect the soundness of the definitions in the present paper, nor does introduce circularities in the definition of abstract equality.

\subsubsection{Model classes.}
The model of a class $C$ is expressed as a collection $D_C = D_C^1, D_C^2, \linebreak \ldots, D_C^n$ of \emph{model classes}.\footnote{The model may include the same class multiple times}
Model classes are immutable classes designed for specification purposes; essentially, they are wrappers of rigorously defined mathematical entities: elementary sorts such as Booleans, integers, and object references, as well as more complex structures such as sets, bags, relations, maps, and sequences.
The MML library \cite{Schoeller2007} provides a variety of such model classes, equipped with features that correspond to common operations on the mathematical structure they represent, including first-order quantification.
For example, class \lstinline|MML_SET| models sets of elements of homogeneous type; it includes features for operations such as membership and quantification over all elements of the set that satisfy a certain predicate (passed as a function object).

\begin{example}
As we discussed in Example \ref{ex:queues}, a sequence is a suitable model for a queue; it can be represented by class \lstinline|MML_SEQUENCE|.
To represent the model of a linked list with internal cursor, we can combine a sequence of class \lstinline|MML_SEQUENCE| with an element of class \lstinline|INTEGER| to represent the position of the cursor; this assumes that no information about the pointer structure of the list in the heap is accessible through the interface of the class.
\end{example}

\subsubsection{Model queries.}
Every class $C$ provides a collection of public \emph{model queries} $S_C = s_C^1, s_C^2, \ldots, s_C^n$, one for each component model class in $D_C$.
Each model query $s_C^i$ returns an instance of the corresponding model class $D_C^i$ that represents the current value of the $i$-th component of the model.
(Informally, the values returned by model queries are analogues to the coefficients expressing the abstract state as a combination of independent basis vectors spanning the whole space).
Since the abstract object state should always be defined between operations and should not depend on the state of any other object, 
model queries are typically argumentless and without precondition.
Clauses in the class invariant can constrain the values of the model queries to match precisely the abstract states of the model.
For example, model query \lstinline|index: INTEGER| returning the cursor position of the \lstinline|LINKED_LIST| in Table \ref{tab:linkedlistetal} should be constrained by an invariant clause \mbox{\lstinline|0 <= index <= sequence.count + 1|}.
A meta-annotation \lstinline|note model: $s_C^1, s_C^2, \ldots$| lists all model queries of the class (see Table \ref{tab:mb-linkedlistetal} for an example).

Programmers can add model queries incrementally to classes developed with DbC.
In fact, it is likely that some model queries are already used in the implementation before models are added explicitly; for example feature \lstinline|index| of class \lstinline|LINKED_LIST| (Table \ref{tab:mb-linkedlistetal}).
Additional model queries return the remaining components of the model for specification purposes, such as \lstinline|sequence| in \lstinline|LINKED_LIST|.

Our approach prefers to implement new model queries as functions rather than attributes.
This choice facilitates a purely descriptive usage of references to model queries in specifications.
In other words, instead of augmenting routine bodies with bookkeeping instructions that update model attributes, routine postconditions are extended with clauses that describe the new value returned by model queries in terms of the old one.
This has the advantage of enforcing a cleaner division between implementation and specification, while better modularizing the latter at routine level (properties of model attributes are typically gathered in the class invariant).
A meta-annotation of the form \lstinline|note specification| tags model queries that are not meant for use in implementation; runtime checking of annotations calling these model queries can be disabled if performance is a concern.

\setcounter{lstnumber}{35}
\lstset{basicstyle=\scriptsize}
\begin{table}[!tb]

\begin{tabular}{m{.46\textwidth} m{.55\textwidth}}
\begin{lstlisting}
note model: sequence, index
class LINKED_LIST [G]
$\ldots$
   has (v: G): BOOLEAN
     -- Does list include `v'? (Reference equality)
     do $\ \ldots$
     ensure Result iff sequence.has (v) end

	item: G
			-- Value at cursor position
		require
			sequence.domain [index]
		ensure
			Result = sequence [index]
		end
\end{lstlisting} \addtocounter{lstnumber}{1}
&
\begin{lstlisting}
  duplicate (n: INTEGER): LINKED_LIST [G]
    -- A copy of at most `n' elements
    -- starting at cursor position
    require n >= 0
    do $\ \ldots$
    ensure
     Result.sequence = sequence.interval (index, index + n - 1)
     Result.index = 0
    end

  make_empty
    -- Create an empty list
    ensure   sequence.is_empty  and  index = 0
    end
$\ldots$
end
\end{lstlisting}
\end{tabular}
\caption{Snippets of class \lstinline|LINKED_LIST| with model-based contracts (continued from Table \ref{tab:mb-linkedlistetal}).}
\label{tab:mb-linkedlistmore}

\end{table}
\lstset{basicstyle=\footnotesize}

\subsubsection{Model-based contracts.}
Let $C$ be a class equipped with model queries whose interface $I_C$ is partitioned into queries $Q_C$ and commands $M_C$.
$Q_C$ now includes the model queries $S_C \subseteq Q_C$ together with other queries $R_C = Q_C \setminus S_C$ 
(note that this does not change the abstract space according to the definitions given at the beginning of the section).
Queries in $R_C$ are called \emph{standard queries}.
The rest of the section contains guidelines to writing model-based contracts for commands in $M_C$ and queries in $R_C$. 

\begin{itemize}
\item The \emph{precondition} of a feature is a constraint on the abstract values of its value-bound arguments and, possibly, on the actual references to its reference-bound arguments.
The target object, in particular, can be considered an implicit value-bound argument.
For example, the precondition \lstinline|map.domain [k]| of feature \lstinline|put| in class \lstinline|TABLE| (Table \ref{tab:mb-linkedlistetal}),
refers to the abstract state of the target object, given by the model query \lstinline|map|,
and to its actual reference-bound argument \lstinline|k|.


\item \emph{Postconditions} should refer to abstract states only through model queries.
This emphasizes the components of the abstract state that a feature modifies or relies upon, which in turn facilitates understanding and reasoning on the semantics of a feature.

\item The \emph{postcondition of a command} defines a relation between the prestate and the poststate of its arguments and the target object; prestate and poststate refer respectively to the state before and after executing the command.
More precisely, the postcondition mentions only abstract values of its value-bound arguments and possibly the actual references to its reference-bound arguments; the target object is considered value-bound both in the prestate and in the poststate.

It is common that a command only affects a few components of the abstract state and leaves all the others unchanged.
Accordingly, the \emph{closed world assumption} is convenient: 
the value of any model query $s \in S_C$ that is not mentioned in the postcondition is assumed not to be modified by the command, 
as if \lstinline|s = old s| were a clause of the postcondition.
When the closed world assumption is wrong, 
explicit clauses in the postcondition should establish the correct semantics.
If a command may modify the value of a model query \lstinline|s| 
but the actual new value is not known precisely and \lstinline|s| is not mentioned in other clauses of the postcondition, 
add a clause \lstinline|relevant (s)| to the postcondition of the command 
(in terms of implementation, \lstinline|relevant| is just a constant function that returns true).
If a command does not affect the value a model query \lstinline|s| 
but the postcondition of the command mentions \lstinline|s|, 
add a clause \lstinline|s = old s| to the postcondition of the command.

\item The \emph{postcondition of a query} defines the result as a function of its arguments and the target object (with the usual discipline of mentioning only abstract values of value-bound arguments and target object and possibly actual references to reference-bound arguments).
Value-bound queries define the abstract state of the result, whereas reference-bound queries describe an actual reference to it.
For example, compare the postcondition of the reference-bound query \lstinline|item| from class \lstinline|LINKED_LIST| (Table \ref{tab:mb-linkedlistmore}), which precisely defines a reference to the returned list element,
with the postcondition of the value-bound query \lstinline|duplicate| in the same class, which specifies the abstract state of the returned list.

\item A clear-cut separation between queries and commands assumes \emph{abstract purity} for all queries: executing a query leaves the abstract state of all its arguments and of the target object unchanged.
\end{itemize}

\setcounter{lstnumber}{2}
\lstset{basicstyle=\scriptsize}
\begin{table}[!tb]

\begin{tabular}{m{.46\textwidth} m{.55\textwidth}}
\begin{lstlisting}
note model: bag
class COLLECTION [G]
  bag: MML_BAG [G]

  is_empty: BOOLEAN
    ensure Result = bag.is_empty end
		
  wipe_out
    ensure bag.is_empty	end

  put (v: G)
    ensure bag = old bag.extended (v) end
end
\end{lstlisting} \addtocounter{lstnumber}{1}
&
\begin{lstlisting}
note model: sequence
class DISPENSER [G]
inherit COLLECTION [G]

  sequence: MML_SEQUENCE [G]

  invariant
    bag.domain = sequence.range
    bag.domain.for_all ( agent (x: G): BOOLEAN 
                   bag [x] = sequence.occurrences (x) )
end
\end{lstlisting}
\end{tabular}
\caption{Snippets of classes \lstinline|COLLECTION| (left) and \lstinline|DISPENSER| (right) with model-based contracts.}
\label{tab:mb-collection-dispenser}
\end{table}
\lstset{basicstyle=\footnotesize}

\subsubsection{Inheritance and model-based contracts.}
A class $C'$ that inherits from a parent class $C$ may or may not re-use $C$'s model queries to represent its own abstract state.
For every model query $s_C \in S_C$ of the parent class that is not among the heir's model queries $S_{C'}$, $C'$ should provide a \emph{linking invariant} to guarantee consistency in the inheritance hierarchy.
The linking invariant is a formula that defines the value returned by $s_C$ in terms of the values returned by the model queries $S_{C'}$ of the inheriting class.
This guarantees that the new model is indeed a specialization of the previous model, in accordance with the notion of sub-typing inheritance.

A properly defined linking invariant ensures that every inherited feature has a definite semantics in terms of the new model.
However, the new semantics may be weaker in that a command whose contract in the parent class characterized it as a function, 
becomes characterized as a relation in the child class; 
that is, incompleteness is introduced (see Section \ref{sec:completeness}).

\begin{example}
Consider class \lstinline|COLLECTION| in Table \ref{tab:mb-collection-dispenser}, a generic container of elements whose model is a bag.
Class \lstinline|DISPENSER| inherits from \lstinline|COLLECTION| and specializes it by introducing a notion of insertion order; correspondingly, its model is a sequence.
The linking invariant of \lstinline|DISPENSER| defines the value of the inherited feature \lstinline|bag| in terms of the new feature \lstinline|sequence|: the domain of \lstinline|bag| coincides with the range of \lstinline|sequence|, and the number of occurrences of any element \lstinline|x| in \lstinline|bag| correspond to the number of occurrences of the same element in \lstinline|sequence|.

The linking invariant ensures that the semantics of features \lstinline|is_empty| and \lstinline|wipe_out| $\!\!$is unambiguously defined also in \lstinline|DISPENSER|.
On the other hand, the model-based contract of command \lstinline|put| in \lstinline|COLLECTION| and the linking invariant are insufficient to characterize the effects of \lstinline|put| in \lstinline|DISPENSER|, as the position within the sequence where the new element is inserted is irrelevant for the bag.
\end{example}

\subsection{Completeness of contracts} \label{sec:completeness}
The notion of \emph{completeness} for the specification of a class gives an indication of how accurate are the contracts of that class with respect to the model.
An incomplete contract does not fully capture the effects of a feature, suggesting that the contract may be more detailed or, less commonly, that the model of the class --- and hence its interface --- is not abstract enough.
Unlike the notion of \emph{sufficient completeness} for algebraic specifications \cite{Guttag1978} --- that serves a similar purpose ---, the present definition of completeness is structurally similar to the concept of completeness for a set of axioms, and a dual notion of soundness complements it.
For simplicity, the following definitions do not mention feature arguments; introducing them is, however, routine.

\subsubsection{Soundness and completeness of a model-based contract.}
Let $f$ be a feature of class $C$.
The specification of $f$ denotes two predicates $\pre_f$ and $\post_f$.
$\pre_f$ represents the set of objects of class $C$ that satisfy the precondition.
If $f$ is a query returning object of class $T$, $\post_f$ has signature $C \times T$ and denotes the pairs of target and returned objects.
If $f$ is a command, $\post_f$ has signature $C \times C$ and denotes the pairs of target objects before and after executing the command.\footnote{These definitions imply the absence of side-effects in evaluating assertions.}

\begin{itemize}
\item The \emph{precondition} of a feature $f$ (query or command) is \emph{sound} iff: for every $o_1, o_2 : C$ such that $o_1 \modeleq_C o_2$ it is $\pre_f(o_1) \Leftrightarrow \pre_f(o_2)$.\footnote{Completeness of preconditions is not an interesting notion and hence it is not defined.}

\item The \emph{postcondition of a command} $m$ is \emph{sound} iff: for every $o, o_1',o_2' : C$ such that $\pre_m(o)$ and $o_1' \modeleq_C o_2'$ it is $\post_m(o,o_1') \Leftrightarrow \post_m(o,o_2')$.

The \emph{postcondition of a command} $m$ is \emph{complete} iff: for every $o, o_1',o_2' : C$ such that $\pre_m(o)$, $\post_m(o,o_1')$, and $\post_m(o,o_2')$ it is $o_1' \modeleq_C o_2'$.

\item The \emph{postcondition of a value-bound query} $q$ is \emph{sound} iff: for every $o : C$ and $t_1, t_2 : T$ such that $\pre_q(o)$ and $t_1 \modeleq_T t_2$ it is $\post_q(o,t_1) \Leftrightarrow \post_q(o,t_2)$.

The \emph{postcondition of a value-bound query} $q$ is \emph{complete} iff: for every $o : C$ and $t_1, t_2 : T$ such that $\pre_q(o)$, $\post_q(o,t_1)$, and $\post_q(o,t_2)$ it is $t_1 \modeleq_T t_2$.

\item The \emph{postcondition of a reference-bound query} $q$ is \emph{sound} iff: for every $o : C$ and $t_1, t_2 : T$ such that $\pre_q(o)$ and $t_1 \equiv_T t_2$ it is $\post_q(o,t_1) \Leftrightarrow \post_q(o,t_2)$.

The \emph{postcondition of a reference-bound query} $q$ is \emph{complete} iff: for every $o : C$ and $t_1, t_2 : T$ such that $\pre_q(o)$, $\post_q(o,t_1)$, and $\post_q(o,t_2)$ it is $t_1 \equiv_T t_2$.
\end{itemize}





Informally, a sound assertion is one that is consistent with the notion of equivalence that is appropriate: 
sound postconditions of commands and value-bound queries do not distinguish between objects with the same abstract state; 
sound postconditions of reference-bound queries do not distinguish between aliases.\footnote{Postconditions of argumentless reference-bound queries are trivially sound for sensible definitions of reference equality.}

A postcondition is complete if all the pairs of objects that satisfy it are equivalent (according to the right model of equivalence).
This means that the complete postcondition of a command defines the effects of the command as a mathematical \emph{function} (as apposed to a relation) 
from the prestate to the abstract poststate.
Similarly, the complete postcondition of a query defines the result as a \emph{function} of the abstract state of value-bound arguments and of actual references to reference-bound arguments.

\begin{example} \label{ex:toincompleteness}
The contracts of features \lstinline|is_empty|, \lstinline|wipe_out|, and \lstinline|put| in class \lstinline|COLLECTION| (Table \ref{tab:mb-collection-dispenser}) are sound and complete; the postcondition of \lstinline|put|, in particular, is complete as it defines the new value of \lstinline|bag| uniquely.
In the heir class \lstinline|DISPENSER|, however, the inherited postcondition of \lstinline|put| becomes incomplete: the linking invariant does not uniquely define \lstinline|sequence| from \lstinline|bag|, hence inequivalent sequences (for example, one with \lstinline|v| inserted at the beginning and another one with \lstinline|v| at the end) satisfy the postcondition.
\end{example}

\subsubsection{Soundness and completeness in practice.}
As the previous example suggests, reasoning informally --- but precisely --- about soundness and completeness of model-based contracts is often straightforward and intuitive, 
especially if the guidelines of Section \ref{sec:spec-class-with} have been followed.
Completeness captures the uniqueness of the (abstract) state described by a postcondition, 
hence query postconditions in the form \lstinline{Result = exp (s, a)} or \lstinline{Result.s = exp (s, a)} 
and command postconditions in the form \lstinline{s = exp (old s, a)} 
--- where \lstinline|exp| is a side-effect free expression, 
\lstinline|s| denotes the value returned by the model query of some argument, and \lstinline|a| is a reference-bound argument 
--- are painless to check for completeness.

\begin{example} 
Consider the following example, from class \lstinline|ARRAY| whose model is a map.
\setcounter{lstnumber}{2}
\begin{lstlisting}
  fill (v: G ; l, u: INTEGER)       -- Put `v' at all positions in [`l', `u'].
   require map.domain [l] and map.domain [u]
   ensure  map.domain = old map.domain
          ( map | {MML_INT_SET} [[l, u]] ).is_constant (v)
          ( map | (map.domain - {MML_INT_SET} [[l, u]]) ) =
                    old ( map | (map.domain - {MML_INT_SET} [[l, u]]) )
   end
\end{lstlisting}
Pre and postconditions are sound because they both refer only to model queries, or functions thereof.
The following reasoning shows that the postcondition is also complete: a map is uniquely defined by its domain and by a value for every key in the domain.
The first clause of the postcondition defined the domain completely.
Then, let $k$ be any key in the domain. If $k \in [l, u]$ then the second clause defines \lstinline|map (k) = v|; otherwise $k \not\in [l, u]$, and the third clause postulates \lstinline|map(k)| unchanged.
\end{example}


Soundness is a mandatory requirement for pre and postconditions in the presence of model-based contracts, as it boils down to writing contracts that are consistent with the chosen level of information hiding.

On the other hand, how useful is completeness in practice?
As a norm, completeness is a valuable yardstick to evaluate whether the contracts are sufficiently detailed.
This is not enough to guarantee that the contracts are correct --- and meet the original requirements --- but the yardstick is serviceable methodologically to focus on what a routine really achieves and how that is related to the abstract model.
As a result, inconsistencies in specifications are less likely to occur, and the impossibility of systematically writing complete contracts is a strong indication that the model is incorrect, or the implementation is faulty.
Either way, a warning is available before attempting a correctness proof.

While complete postconditions should be the norm, there are recurring cases where incomplete postconditions are unavoidable or even preferable.
Three major sources of benign incompleteness are the following.
\begin{itemize}
\item Inherently \emph{nondeterministic or stochastic} specifications.
For example, a class for random number generation can use a sequence as model, but its specification should not define the precise content of the sequence unambiguously.

\item Usage of \emph{inheritance} to factor out common parts of (complete) specifications.
For example, class \lstinline|DISPENSER| in Table \ref{tab:mb-collection-dispenser} is a common ancestor of \lstinline|STACK| and \lstinline|QUEUE|.
If its interface includes features \lstinline|item|, \lstinline|put| and \lstinline|remove|, its model must be isomorphic to a sequence.
Then, it becomes impossible to write a complete postcondition for \lstinline|put| in \lstinline|DISPENSER|: 
the specification of \lstinline|put| cannot define precisely where an element is added to the sequence;
a choice compatible with the semantics of \lstinline|STACK| will be incompatible with \lstinline|QUEUE| and vice versa.
 
\item Imperfections in \emph{information hiding}.
For example, class \lstinline|ARRAYED_LIST| is an array-based implementation of lists which exports a query \lstinline|capacity| returning the size of the underlying array; this piece of information is then part of the model of the class.
Default constructors set \lstinline|capacity| to an initial fixed value.
Their postconditions, however, do not mention this default value, hence they are incomplete.
The rationale behind not revealing this information is that clients should not rely on the exact size of the array when they invoke the constructor.
\end{itemize}
In all these cases, reasoning about completeness is still likely to improve the understanding of the classes and to question constructively the choices made for interfaces and inheritance hierarchies.


\subsection{Verification: proofs and runtime checking}
This subsection outlines the main ideas behind using model-based contracts for verification with formal correctness proofs and with runtime checking for automated testing.
Its goal is not to detail any particular proof or testing technique, but rather to sketch how to express the semantics of model-based contracts within standard verification frameworks.


\subsubsection{Proofs.}
The \emph{axiomatic} treatment of model classes~\cite{Charles2006,Schoeller2007,Darvas2007} is quite natural: the semantics of a model class is defined directly in terms of a theory expressed in the underlying proof language, rather than with ``special'' contracts.
The mapping is often straightforward, and has the advantage of reusing theories that are optimized for effective usage with the proof engine of choice.
In addition, the immutability (and value semantics) of model classes makes them very similar to mathematical structures and facilitates a straightforward translation into mathematical theories.

\setcounter{lstnumber}{2}
\lstset{basicstyle=\scriptsize}
\begin{table}[!tb]
\begin{tabular}{m{.48\textwidth} m{.52\textwidth}}
\begin{lstlisting}
note mapped_to: "Sequence G"
class	MML_SEQUENCE [G]
$\ \ldots$
  extended (x: G): MML_SEQUENCE[G]
    -- Current sequence extended with `x' at the end
    note mapped_to: "Sequence.extended(Current, x)"
    do ... end
end
\end{lstlisting} \addtocounter{lstnumber}{1}
&
\hbox{\lstset{language=BoogiePL,firstnumber=9,numbers=left}
\begin{lstlisting}
type Sequence T = [int] T ;
function Sequence.extended $\langle$T$\rangle$ (Sequence T, T)
    returns (Sequence T);
axiom (forall $\langle$T$\rangle$ s: Sequence T, x:T ::{Sequence.extended(s,x)} 
  Sequence.extended(s, x) == s[Sequence.count(s)+1 := x]);
axiom (forall $\langle$T$\rangle$ s: Sequence T, x: T ::
         {Sequence.count(Sequence.extended(s, x))}
  Sequence.count(Sequence.extended(s, x)) ==
           Sequence.count(s)+1);
$\ \ldots$
\end{lstlisting}}
\end{tabular}
\caption{Snippets from class \lstinline|MML_SEQUENCE| (left) and the corresponding Boogie theory (right).}
\label{tab:mmlsequence}
\end{table}
\lstset{basicstyle=\footnotesize}

In this respect, we are currently developing an accurate mapping of model classes and model-based contracts into Boogie~\cite{Specsharp}.
First, the mapping introduces axiomatic definitions of MML model classes as Boogie theories; annotations in the form \lstinline|note mapped_to| connect MML classes to the corresponding Boogie types.
For example, Table \ref{tab:mmlsequence} shows how a portion of the \lstinline{MML_SEQUENCE} model class translates into a Boogie theory: a mapping type \lstinline[language=BoogiePL]|[int] T| represents sequences of elements of generic type \lstinline[language=BoogiePL]|T|, and a few axioms constrain a function \lstinline[language=BoogiePL]|Sequence.extended| to return values in accordance with the MML semantic of feature \lstinline|extended|.

Then, each model query in a class with model-based contracts maps to a Boogie function that references a representation of the heap; some axioms connect the value returned by the function to other features in the translated class.
For example, the model query \lstinline|sequence| in \lstinline|LINKED_LIST| becomes \lstinline[language=BoogiePL]|function LinkedList.sequence(HeapType, ref) returns (Sequence ref)|.



Finally, model-based contracts are translated into Boogie formulas according to the \lstinline{mapped_to} annotations in model classes.
For example, the postcondition clause:\linebreak
\lstinline|sequence = old (sequence.front (index).extended (v) + sequence.tail (index + 1))|
of \lstinline|put_right| in \linebreak \lstinline|LINKED_LIST| (Table \ref{tab:mb-linkedlistetal}) maps to the Boogie formula:
\begin{lstlisting}[language=BoogiePL]
LinkedList.sequence(Heap, Current) ==  Sequence.concat ( Sequence.extended (
     Sequence.front(LinkedList.sequence(old(Heap), Current),
                    LinkedList.index(old(Heap), Current)),   v ),
     Sequence.tail(LinkedList.sequence(old(Heap), Current), 
                   LinkedList.index(old(Heap), Current) + 1) );
\end{lstlisting}

\subsubsection{Runtime checking and testing.}
Most model classes represent \emph{finite} mathematical objects, such as sets of finite cardinality, sequences of finite length, and so on.
All these classes can have an implementation of their operations which is executable in finite time; this supports the runtime checking of assertions that reference these model classes.

Testing techniques can leverage runtime checkable contracts to fully automate the testing process: generate objects by randomly calling constructors and commands; check the precondition of a routine on the generated objects to filter out valid inputs for the routine; execute the routine body on a valid input and check the validity of the postcondition on the result; any postcondition violation on a valid input is a fault in the routine.

This approach to contract-based testing has proved extremely effective at uncovering plenty of bugs in production code \cite{Meyer2009}, hence it is an excellent ``lightweight'' precursor to correctness proofs.
Contract-based testing, however, is only as good as the contracts are; the weak postconditions of traditional DbC, in particular, leave many real faults undetected.
Runtime checkable model-base contracts can help in this respect and boost the effectiveness of contract-based testing by providing more expressive, and complete, specifications.
Section \ref{sec:mbc-at-work} describes some testing experiments that support this claim.

\subsubsection{Consistency of tests and proofs.}
Using contract-based testing as a precursor to correctness proofs poses the problem of consistency between two semantics given to model classes: the runtime semantics given by an executable implementation and the proof semantics given by a mapping to a logical theory.
Under reasonable assumptions about the execution environment, consistency must ensure that a component is proven correct against its model-based specification if and only if testing the component never detects a violation of its model-based contracts.
Establishing this consistency amounts to proving that: (1) the implementation of each model class is consistent with the mapping of the class to a logical theory; and (2) the implementation of each model query satisfies its specification.
Future work will detail and address these problems.

\section{Model-based contracts at work}\label{sec:mbc-at-work}
This section describes experiments in developing model-based contracts for real object-oriented software written in Eiffel.
The experiments target two non-trivial case studies based on data-structure libraries (described in Section \ref{sec:case-studies}) with the goal of demonstrating that deploying model-based contracts is feasible, practical, and useful.
Section \ref{sec:results-and_discussion} discusses the successes and limitations highlighted by the experiments.

\subsection{Case studies}  \label{sec:case-studies}
The first case study targeted EiffelBase \cite{EiffelBase}, a library of general-purpose data structures widely used in Eiffel programs; EiffelBase is representative of mature Eiffel code exploiting extensively traditional DbC.
We selected 7 classes from EiffelBase, for a total of 304 features (254 of them are public) over more that 5700 lines of code.
The 7 classes include 3 widely used container data structures (\lstinline|ARRAY|, \lstinline|ARRAYED_LIST|, and \lstinline|LINKED_LIST|)
and 4 auxiliary classes used by the containers (\lstinline|INTEGER_INTERVAL|, \lstinline|LINKABLE|, \lstinline|ARRAYED_LIST_CURSOR|, and \lstinline|LINKED_LIST_CURSOR|).
Our experiments systematically introduced models and conservatively augmented the contracts of all public features in these 7 classes with model-based specifications.

The second case study developed EiffelBase2, a new general-purpose data structure library.
The design of EiffelBase2 is similar to that of its precursor EiffelBase; EiffelBase2, however, has been developed from the start with expressive model-based specifications and with the ultimate goal of proving its full functional correctness --- backward compatibility is not one of its primary aims.
This implies that EiffelBase2 rediscusses and solves any deficiency and inconsistency in the design of EiffelBase that impedes achieving full functional correctness or hinders the full-fledged application of formal techniques.
EiffelBase2 provides containers such as arrays, lists, sets, tables, stacks, queues, and binary trees; iterators to traverse these containers; and comparator objects to parametrize containers with respect to arbitrary equivalence and order relations on their elements.
The current version of EiffelBase2 includes 46 classes with 460 features (403 of them are public) totaling about 5800 lines of code; these figures make EiffelBase2 a library of substantial size with realistic functionalities.
The latest version of EiffelBase2 is available at \url{http://eiffelbase2.origo.ethz.ch}.

\subsection{Results and discussion}  \label{sec:results-and_discussion}
This section addresses the following questions based on the experience with the two case studies of EiffelBase and EiffelBase2.
\begin{itemize}
\item How many different model classes are needed to write model-based contracts?
\item How many contracts can be complete?
\item Do executable accurate model-based contracts boost contract-based testing?
\end{itemize}

\setcounter{lstnumber}{2}
\lstset{basicstyle=\scriptsize}
\begin{table}[!tb]
\begin{tabular}{m{.48\textwidth} m{.55\textwidth}}
\begin{lstlisting}
note model: set, relation
class SET [G]
$\ \ldots$
  has (v: G): BOOLEAN
    -- Does this set contain `v'?
    ensure
     Result = not (set * relation.image_of (v)).is_empty
    end

  set: MML_SET [G] -- The set of elements
  relation: MML_RELATION [G, G]
			-- Equivalence relation on elements
end
\end{lstlisting}  \addtocounter{lstnumber}{1}
&
\begin{lstlisting}
note model: map
class BINARY_TREE [G]
$\ \ldots$
  add_root (v: G)
    -- Add a root with value `v' to an empty tree
    require   map.is_empty
    ensure    map.count = 1  and  map [Empty] = v
    end

  map: MML_MAP [MML_SEQUENCE[BOOLEAN], G]
    -- Map of paths to elements
end
\end{lstlisting}
\end{tabular}
\caption{Examples of nonobvious models: classes \lstinline|SET| and \lstinline|BINARY_TREE| from EiffelBase2.}
\label{tab:nontrivial}

\end{table}
\lstset{basicstyle=\footnotesize}

\subsubsection{How many model classes?}
Model-based contracts for EiffelBase used model classes for Booleans, integers, references, (finite) sets, relations, and sequences.
EiffelBase2 additionally required (finite) maps, bags, and infinite maps and relations for special purposes (such as modeling comparator objects).
These figures suggest that a moderate number of well-understood mathematical models suffices to specify a general-purpose library of data structures.

Determining to what extent this is generalizable to software other than libraries of general-purpose data structures is an open question which belongs to future work.
Domain-specific software may indeed require complex domain-specific model classes (e.g., real-valued functions, stochastic variables, finite-state machines), and application software that interacts with a complex environment may be less prone to accurate documentation with models.
However, even if writing model-based contracts for such systems proved exceedingly complex, some formal model is required if the goal is formal verification.
In this sense, focusing model-based contracts on library software is likely to have a great payoff through extensive reuse: the many clients of the reusable components can rely on expressive contracts not only as detailed documentation but also to express their own contracts and interfaces by combining a limited set of well-understood, highly dependable components.

Another interesting remark is that the correspondence between the limited number of model classes needed in our experiments and the classes using these model classes is far from trivial: data structures are often more complex than the mathematical structures they implement.
Consider, for example, class \lstinline|SET| in Table \ref{tab:nontrivial}: EiffelBase2 sets are parameterized with respect to an equivalence relation, hence the model of \lstinline|SET| is a pair of a mathematical set and a relation.
Another significant example is \lstinline|BINARY_TREE| (also in Table \ref{tab:nontrivial}): instead of introducing a new model class for trees or graphs, \lstinline|BINARY_TREE| concisely represents a tree as a map of paths to values; the model of a path is in turn a sequence of Booleans.

\subsubsection{How many complete contracts?}
Reasoning informally, but rigorously, about the completeness of postconditions --- along the lines of Section \ref{sec:completeness} --- proved to be straightforward in our experiments.
Only 18 (7\%) out of 254 public features in EiffelBase with model-based contracts and 17 (4\%) out of 403 public features in EiffelBase2 have incomplete postconditions.
All of them are examples of ``intrinsic'' incompleteness mentioned at the end of Section \ref{sec:completeness}; EiffelBase2, in particular, was designed trying to minimize the number of features with intrinsically incomplete postconditions.

These results indicate that model-based contracts make it feasible to write systematically complete contracts; in most cases this was even relatively straightforward to achieve.
Unsurprisingly, using model-based contracts dramatically increases the completeness of contracts in comparison with standard DbC.
For example, 42 (66\%) out of 64 public features of class \lstinline|LIST| in the original version of EiffelBase (without model-based contracts) have incomplete postconditions, including 20 features (31\%) without any postcondition.

\setcounter{lstnumber}{2}
\lstset{basicstyle=\scriptsize}
\begin{table}[!tb]
\begin{lstlisting}
merge_right (other: LINKED_LIST [G])
  -- Merge `other' into current list after cursor position. Do not move cursor. Empty `other'.
  do
    $\ldots\ $
    other_first_element := other.first_element  ;  other_count := other.count  ;  other.wipe_out
    if before then  first_element := other_first_element  ;  active := first_element
    else $\ \ldots\ $ end
    count := count + other_count
  ensure
    -- Original contract
    count = old count + old other.count  ;  index = old index  ;  other.is_empty
	 -- Model based contract
	 sequence = old (sequence.front (index) + other.sequence + sequence.tail (index + 1))
  end
\end{lstlisting}
\caption{Faulty routine \lstinline|merge_right| from class \lstinline|LINKED_LIST|.}
\label{tab:merge}
\end{table}

\lstset{basicstyle=\footnotesize}

\subsubsection{Contract-based testing with model-based contracts.}
The standard EiffelBase library has been in use for many years and has been extensively tested, both manually and automatically.
Are the expressive contracts based on models enough to boost automated testing finding new, subtle bugs?
While preliminary, our experiments seem to answer in the affirmative.
Applying the AutoTest testing framework \cite{Meyer2009} on EiffelBase with model-based contracts for 30 minutes discovered 3 faults; none of them would have been detectable with standard contracts.
Running these tests did not require any modification to AutoTest or model classes, because the latter include an executable implementation.

The 3 faults reveal subtle mistakes that have gone undetected so far.
For example, consider the implementation of routine \lstinline|merge_right| in Table \ref{tab:merge}; the routine merges a linked list \lstinline|other| into the current linked list at the cursor position by modifying references in the chain of elements.
The \lstinline|then| branch of the \lstinline|if| statement (line 6) deals with the special case where the cursor in the current list is \lstinline|before| the first element; in this case the first element of the current list (\lstinline|first_element|) will point directly to the first element of the other list (\lstinline|other_first_element|).
This is not sufficient, as the routine should also link the end of the other list to the front of the current one, otherwise all elements in the current list become inaccessible.
The original contract does not detect this fault; the clause \lstinline|count = old count + old other.count| is in particular satisfied as \lstinline|count| is updated anyway (line 8), but its value does not reflect the actual content of the new list.
On the contrary, the complete model-based contract (line 13) specifies the desired configuration of the list after executing the command, which leads to easily detecting the error.


\section{Related work} \label{sec:related-work}
Every fully formal specification ultimately boils down to a mathematical model, and the research on formal modeling and analysis is so extensive and diverse that it cannot be summarized concisely.
This section focuses on a few major approaches to the formal specification of object-oriented abstract data types that adopt a stance similar to that of the present paper: using highly expressive mathematical models geared towards the full functional correctness specification (and verification) of complex data structures.

Hoare pioneered the usage of mathematical models to define and prove correctness of data type implementations \cite{Hoare72}.
This idea spawned much related work, which can be roughly partitioned in three major lines: algebraic notations, descriptive notations, and design-by-contract approaches.
The following subsections shortly summarize the main features of each of these techniques; then, Section \ref{sec:model-based-annot} describes the approaches based on mathematical models that are closest to the present paper.

\subsection{Algebraic notations}
Algebraic notations formalize classes in terms of (uninterpreted) functions and axioms that describe the mutual relationship among the functions. 
For example, the axiom $s.\SETinsert(x).\SETmemberof(x)= \logictrue$ defines the mutual semantics of the operations $\SETinsert$ and $\SETmemberof$ of a set data type.
The most influential work in algebraic specifications is arguably Guttag and Horning's \cite{Guttag1978} and Gougen et al.'s \cite{gtw78}, which gave a foundation to much derivative work.
The former was also made practical in the Larch project \cite{Guttag1993}, and introduced a notion of \emph{completeness} that differs from the one of the present paper (see Section \ref{sec:completeness}), and applies to whole types, not single features.

Algebraic notations emphasize the calculational aspect of a specification.
This makes them very effective notations to formalize and verify data types at a high level of abstraction.
In particular, the close connection between rewriting systems \cite{DJ90} and algebraic definitions enables, in many practical cases, the automated or semi-automated verification of consistency and completeness \cite{Guttag1978} requirements of abstract specifications.
The algebraic approach, on the other hand, does not integrate as well with real programming languages to document implementations in the form of pre and postconditions of single operations.

\subsection{Descriptive notations}
Descriptive notations formalize classes in terms of simpler types --- ultimately grounded in simple mathematical models such as sets and relations --- and operations defined as input/output relations (that is, pre and postconditions) constrained by logic or arithmetic formulas.
For example, the $\SETinsert$ operation of a set data structure could be defined by the formula $\forall s, x \bullet \ldoubsq s.\SETinsert(x) \rdoubsq = \ldoubsq s \rdoubsq \cup \{ x \}$, in terms of the union operation applied to a model set $\ldoubsq s \rdoubsq$.

Descriptive notations can be used in isolation to build language-independent models, or to give a formal semantics to concrete implementations.
Languages and methods such as Z \cite{Woodcock1996}, B \cite{Abrial1996}, and VDM \cite{Jones1990} pursue the former approach, usually within a top-down development framework.
Other specification languages and tools such as RESOLVE \cite{RESOLVE}, AAL \cite{AAL}, and Jahob \cite{Jahob} are examples of the latter approach for the programming languages C$^{++}$ and Java.

Descriptive notations are apt to develop correct-by-construction designs and to accurately document implementations, often with the goal of verifying functional correctness.
Using them in contracts, however, introduces a new notation on top of the programming language, which requires additional effort and expertise from the programmer and makes it more difficult to to maintain the specification synchronized with the actual implementation.
This weakness is shared by algebraic notations alike.

\subsection{Design-by-contract approaches}
Design by contract \cite{OOSC2} introduces formal specifications in programs using the same notation for implementation and annotations, in an attempt to make writing the contracts as congenial as possible to programmers.
The Eiffel programming language \cite{ECMAstandard} epitomizes the design by contract methodology, together with similar solutions for other languages such as APP \cite{APP} for C, Spec$^\#$ \cite{Specsharp} for C$^\#$, and many others.

As we discussed also in the rest of the paper, using a subset of the programming language in annotations helps programmers writing them \cite{WritingContracts}, but it often does not provide enough expressive power to formalize (easily) ``complete'' functional correctness, or requires cumbersome workarounds to capture the semantics of mathematical concepts in terms of programming language constructs.

\subsection{Model-based annotation languages} \label{sec:model-based-annot}
The Java Modeling Language (JML) \cite{Leavens2005,Leavens2006} is likely the approach that shares the most similarities with ours: JML annotations are based on a subset of the Java programming language and the JML framework provides a library of model classes mapping mathematical concepts.
While sharing a common outlook, the approaches in JML and in the present paper differ in several details pertaining scope and technical aspects.

At the technical level, JML prefers model variables \cite{Cheon2005} while our approach leverages model queries that return the value of immutable model classes; each approach has its merits, but model queries have the advantage of supporting an axiomatic definition that is easily grounded in an underlying mathematical theory, and facilitate a seamless integration with traditional contracts --- also typically based on queries.
Section \ref{sec:spec-class-with} discusses other advantages of model queries.
A notational difference is that JML extends Java's expressions with notations for logic operators and quantifiers, while our method does not extend Eiffel's syntax and reuses notation such as agents to express quantifications and other aspects that belong to expressive specifications.

In terms of scope, our approach strives to be more methodological and systematic, with the primary target of fully contracting a complete library of data structures.
Our method tries to keep the additional effort required to the programmer to a minimum. 
Finally, let us remark that our usage scenarios are multi-faceted, ranging from specification and design (also supporting notions such as completeness), to verification, runtime
 checking, and automated testing.

The present paper extends in scope the previous work of ours on model-based classes \cite{Schoeller2004,Schoeller2007}, and systematically applies the results to the re-design and re-im\-ple\-men\-ta\-tion of a rich library of data structures.
The experience gained in this practical application also prompted us to refine and rediscuss aspects of the previous approach, as we discussed at length in the rest of the paper.


\section{Conclusions and future work} \label{sec:conclusions}
The present work introduces a methodology to write strong interface specifications for reusable object-oriented components.
The methodology is soundly based on expressive models based on mathematical notions and features a notion of specification completeness which is formal, yet easy to reason about.
The application of the methodology to the development of a library of general-purpose data structures demonstrates its practicality and its many uses in analysis, design, and verification.

Future work includes short- and long-term goals.
Among the former, we plan to apply model-based contracts to more real-life examples, including application software from diverse domains.
A user study will try to confirm the preliminary evidence that model-based contracts are easy to write, understand, and reason about informally.

Longer term work will integrate model-based contracts within a comprehensive verification environment.
This will require, in particular, significant developments in the techniques for proofs and tests with model-based contracts.
Work on proofs will include dealing systematically with the frame problem and extensions of the model-based contract methodology to non-public features, including abstraction functions, representation invariants, and loop invariants.
Work on testing will focus on optimizing the runtime performance of model classes.






\bibliographystyle{abbrv}
\bibliography{mbc_bibliography,mml_relatedwork}
\end{document}